\documentclass[12pt]{iopart}
\usepackage{graphicx}

\begin{document}

\title{Redfield Reduced Dynamics and Entanglement}

\author{Sebastiano Anderloni$^{a,b}$, Fabio Benatti$^{a,b}$, Roberto Floreanini$^{b}$}
\address{$^a$Dipartimento di Fisica Teorica, Universit\`a di Trieste,
Strada Costiera 11,\\
34014 Trieste, Italy\\
${}^b$Istituto Nazionale di Fisica Nucleare, Sezione di Trieste, Strada Costiera 11,\\
34014 Trieste, Italy}

\begin{abstract}
In phenomenological applications the time evolution of subsystems immersed in an external environment are
sometimes described by Markovian semigroups of Redfield type that result non-positive: 
the appearence of negative probabilities is avoided by restricting the admissible initial conditions
to those states that remain positive under the action of the dynamics. We show
that this often adopted procedure may lead to physical inconsistencies in presence of
entanglement.

\end{abstract}

\maketitle

\section{Introduction}

The dissipative evolution of a system $S$ immersed in a noisy environment $E$ 
can be described via dynamical semigroups of linear maps 
$\gamma_t$, acting on density matrices $\rho$ representing the states of the system. 
This reduced dynamics for $S$ alone is obtained from the unitary time evolution
of the full system $S+E$ by tracing over the environment degrees of freedom and
by further adopting a Markovian ({\it i.e.} memoryless) approximation. This procedure
is physically justified when the interaction between $S$ and $E$ is weak and it has
been successfully used in many phenomenological applications in quantum chemistry,
quantum optics and atomic physics [1-12].

Nevertheless, the derivation of such reduced dynamics from the microscopic subsystem-environment
interaction is often based on {\it ad hoc}, simplifying assumptions. 
As a consequence, the resulting 
reduced time evolutions are of Bloch-Redfield type [7, 8, 5]
and therefore might not be fully consistent: 
typically, such naive dynamics do not preserve the positivity
of the $S$ density matrix.%
\footnote{Exceptions to this general result are obtained
using rigorous mathematical treatments [1-5].}

In order to cure these inconsistencies, the general attitude is to restrict the action of 
the non-positive dynamics to a subset of all possible initial density matrices,
a procedure sometimes referred to as ``slippage of the initial conditions'' [13-16].
Physically, this prescription is ascribed to the short-time correlations
in the environment, that are usually neglected in
the derivation of the Markovian limit.

In the following, we shall critically examine this prescription to cure
inconsistencies produced by non-positive, Markovian evolutions 
and point out potential problems of this approach, in particular
in presence of entanglement. More specifically, we shall study the
behaviour of two subsystems, one immersed in the environment and
evolving with a Redfield type dissipative dynamics, while the other
does not evolve at all. We shall see that when the initial
state of the two subsystems is entangled, the ``slippage'' prescription 
does not cure all possible inconsistencies of the two-subsystem dynamics.

Preliminary, partial results on this line of 
investigation have been reported in Refs.[17-19].
In particular, in Ref.[18] the non-positive, 
dissipative evolution has been chosen in an {\it ad hoc}
and rather abstract way in order to expose the just mentioned difficulties.
Here, as reduced dynamics we adopt instead a Redfield non-positive evolution
$\gamma_t$ that has been used in various phenomenological applications [7-12].
With the help of both analytic and numerical methods, we shall then analyze the fate
of the quantum correlations of two qubit systems when they evolve
with a factorized dynamics $\gamma_t\otimes {\rm id}$:
the ``slippage'' prescription is at work for the first qubit, while the other
is inert and evolves with 
the identity operator. We shall explicitly show
that in such a situation the purely local evolution $\gamma_t\otimes {\rm id}$
can increase
the entanglement of the two systems, a clearly unphysical result.
Therefore in presence of entanglement, 
the above mentioned prescription of restricting initial conditions
to cure non-positive, Markovian dynamics does not seem completely satisfactory.

\section{Single system dissipative dynamics}

We shall first study the dynamics of a single subsystem immersed in an external environment.
As explained in the introductory remarks, the physical system we shall consider 
is a two-level system (qubit) immersed in a thermal bath. 
The system is described by $2\times2$ density matrices $\rho$, 
{\it i.e.} by positive Hermitian operators, with unit trace.
On the other hand, the bath is modeled as an infinite dimensional reservoir in equilibrium 
at temperature $T\equiv\beta^{-1}$. Being infinitely large, the environment can be considered
unaffected by the interaction with the subsystem and therefore to be in the reference equilibrium state
\begin{equation}
\rho_{E}=\frac{e^{-\beta H_{E}}}{Tr(e^{-\beta H_{E}})}\ ,
\label{1}
\end{equation}
where $H_{E}$ is the hamiltonian describing the free dynamics of the bath.

The total Hamiltonian describing the evolution 
of the compound system can be chosen as [8, 14-16]
\begin{equation}
H=H_{S}\otimes {\bf 1}_E+{\bf 1}_S\otimes H_{E}+\lambda H_{I}\ ,
\label{2}
\end{equation}
where
\begin{equation}
H_S=\frac{\omega}{2}\sigma_{3}\ ,\qquad\qquad H_I=\sigma_{1}\otimes B\ ,
\label{3}
\end{equation}
are respectively the subsystem Hamiltonian and interaction term, the parameter $\lambda$
playing the role of an adimensional coupling constant; 
$\sigma_i$, $i=1,2,3$ are the Pauli matrices and represent the subsystem operators,
while $B$ is an environment operator, taken for simplicity to satisfy the condition
$Tr_{E}(\rho_{E}B)=0$.

Using standard second order approximation in the coupling constant $\lambda$
and a {\it naive} Markovian limit, one finds that the time evolution
of the reduced density matrix $\rho$ for the system $S$ is generated by a master
equation of Bloch-Redfield type that takes the explicit form [7-9, 5]:
\begin{equation}
{\partial\rho(t)\over \partial t}=
-i\big[H_S ,\rho(t)\big] + L[\rho(t)]\ .
\label{4}
\end{equation}
Besides the standard hamiltonian piece, the r.h.s. contains the extra contribution $L$,
a linear map representing the effects of noise induced by the presence of the external bath.
By setting,
\begin{equation}
B(t)=e^{itH_E}\, B\, e^{-itH_E}\ ,
\label{5}
\end{equation}
and further introducing the environment two-point correlation functions,
\begin{equation}
G(t)={\rm Tr}\big[\rho_E B(t) B\big]={\rm Tr}\big[\rho_E B B(-t)\big]\ ,
\label{6}
\end{equation}
it can explicitly be written as [5]:
\begin{eqnarray}
\nonumber
\hskip -1cm
L[\rho(t)]=&&\lambda^2\int_{0}^{\infty}ds \bigg\{G(s)\Big[\cos(\omega s)[\sigma_{1},\sigma_{1}\rho(t)]
-\sin(\omega s)[\sigma_{2},\sigma_{1}\rho(t)]\Big] \\
&&\hskip 2cm +G(-s)\Big[\cos(\omega s)[\rho(t)\sigma_{1},\sigma_{1}]
-\sin(\omega s)[\rho(t)\sigma_{1},\sigma_{2}]\Big]\bigg\}\ .
\label{7}
\end{eqnarray}
Therefore, the effects of dissipation and noise can be conveniently parametrized in terms of
the following three phenomenological constants:
\begin{eqnarray}
\label{8}
\nonumber
a=\lambda^{2}\int_{0}^{\infty} d s\cos(\omega s)\big[G(s)+G(-s)\big]\ ,\\
b=\lambda^{2}\int_{0}^{\infty} d s\sin(\omega s)\big[G(s)+G(-s)\big]\ ,\\
\nonumber
d=i\lambda^{2}\int_{0}^{\infty} d s\sin(\omega s)\big[G(s)-G(-s)\big]\ .
\end{eqnarray}
Note that these parameters are not completely arbitrary. Indeed, since
$\rho_{E}$ is a thermal state, 
it obeys the so called \textit{Kubo-Martin-Schwinger condition} \cite{20}:
\begin{equation}
G(t)=G(-t-i\beta)\ .
\label{9}
\end{equation}
As a consequence, the parameters $a$ and $d$ above obey the following
relation:
\begin{equation}
a-d=e^{-\beta\omega}(a+d)\ .
\label{10}
\end{equation}
Further, one can show that the coefficient $a$ must be positive [5].

The time evolution of the entries of the density matrix $\rho$,
\begin{equation}
\rho=\left(\matrix{
\rho_1 & \rho_3\cr
\rho_3^* & \rho_2\cr
}\right)\ ,
\label{11}
\end{equation}
can now be explicitly given in terms of the parameters $\omega$, $a$, $b$ and $d$:
\begin{eqnarray}
\label{12}
\hskip -1cm
\nonumber
\rho_1(t)={1\over2}\bigg(1-{d\over a}\bigg)\Big(1-e^{-2at}\Big)+\rho_1(0)\, e^{-2at}\ ,\\
\hskip -1cm
\rho_2(t)=1-\rho_1(t)\ ,\\
\hskip -1cm
\nonumber
\rho_3(t)=e^{-at}\bigg\{\bigg( \cos(\Omega t)
-i{(\omega+b)\over\Omega}\sin(\Omega t)\bigg)\rho_3(0)
+{(a+ib)\over\Omega}\sin(\Omega t)\, \rho_3^*(0)\bigg\}\ ,
\end{eqnarray}
where $\Omega=[\omega^{2}+2b\omega-a^{2}]^{1/2}$.

Unfortunately, this evolution does not 
preserve the positivity of the eigenvalues of $\rho$ for all times.
In order to show this, it is sufficient to consider 
the following initial state $\hat\rho$ with entries:
\begin{equation}
\hat\rho_1={1\over2}\bigg(1-{d\over2a}\bigg)\ ,\qquad 
\hat\rho_3={1\over4}\bigg(1+i{b\over a}\bigg)\, \sqrt{{4a^2-d^2\over a^2+b^2}}\ .
\label{13}
\end{equation}
It is a pure state, since ${\rm Det}[\hat\rho]=\,0$. To have a $\hat\rho$ with positive spectrum,
its determinant must remain non-negative
for all times; in particular, its time derivative at $t=\,0$ must be positive, otherwise
${\rm Det}[\hat\rho]$ would assume negative values as soon as $t>0$. On the other hand,
using (\ref{13}), one easily sees that:
\begin{equation}
\frac{d}{dt} {\rm Det}(\hat\rho)\big|_{t=0}=-\frac{a(4b^{2}+d^{2})}{4(a^{2}+b^{2})}<0\ ,
\label{14}
\end{equation}
a zero value being allowed only when $b=d=0$. Further, note that because of the KMS
condition (\ref{10}), even if $b=\,0$, a vanishing $d$ can be obtained only at infinite temperature,
{\it i.e.} when $\beta=\,0$. It thus follows that at finite temperature the Markov approximation
leading to the master equation (\ref{4}), with generator as in (\ref{7}), 
does not preserve positivity, since a state like $\hat\rho$
is immediately turned into a matrix with negative eigenvalues as soon as $t>0$.

Although unphysical, the time evolution (\ref{12}) generated by (\ref{4}) is nevertheless used in
phenomenological applications because of its good asymptotic behaviour. In fact, it possesses
a unique equilibrium state $\rho_{\rm eq}$, that can be easily determined by setting to zero
the r.h.s. of (\ref{4}):
\begin{equation}
\rho_{\rm eq}=
\frac{1}{{\rm e}^{\beta\omega/2}+{\rm e}^{-\beta\omega/2}}
\pmatrix{
{\rm e}^{-\beta\omega/2}&0\cr 0&{\rm e}^{\beta\omega/2}
}
=\frac{{\rm e}^{-\beta H_S}}{{\rm Tr}[{\rm e}^{-\beta H_S}]}\ .
\label{15}
\end{equation}
Therefore, it turns out that for asymptotically 
long times the system is driven to a thermal state at the bath temperature, 
a behaviour which is physically expected in such open quantum systems.

In order to adopt the Redfield dynamics given in (\ref{12}) as a {\it bona fide} time evolution,
a cure to the non-positivity needs to be introduced. The general solution that has been proposed
is to restrict the space of initial conditions to those states $\rho(0)$ that remain positive
under the action of the Redfield dynamics. The general argument supporting this choice is that
any Markovian approximation neglects a certain initial span of time, the transient, during
which memory effects can not be ignored. During this short transient time, 
the environment acts in a very complicated way on the subsystem and the net result is the
elimination of all states, like $\hat\rho$ in (\ref{13}), that would give rise to inconsistencies
during the subsequent Markovian regime. This mechanism is known in the literature
as ``slippage of initial conditions'' [13-16]. As we shall see in the next section, this
prescription may cure the positivity preserving problem for a single subsystem,
but appears to be inconclusive when dealing with bi- or multi-partite
open systems in view of the existence of entangled states.

\section{Two qubit dynamics and entanglement}

We shall now extend the treatment discussed so far to the case of two qubits, one of which is
still immersed in a heat bath and therefore evolves with the dissipative dynamics $\gamma_t$
generated by the Redfield equations (\ref{4}), while the other remains inert (it is usually
called an {\it ancilla}). The total time evolution for the two qubits
is then in factorized form, $\gamma_t\otimes {\rm id}$, where ``id'' is
the identity operator acting on the second qubit.
In order to have a consistent time evolution, we shall further
assume the ``slippage prescription'' at work for the first qubit:
we remark that this prescription originates in the action of the bath during the 
transient and therefore can only involve the qubit inside the bath, and not the ancilla.

Within this framework, we shall explicitly show the existence of states for the two qubits
that 1) when traced over the ancilla degrees of freedom, belong
to the set of admissible initial states for the
non-positive dynamics $\gamma_t$, 2) remain positive under 
the action of the extended dynamics $\gamma_t\otimes {\rm id}$
and 3) nevertheless present an increase of their entanglement. 
This is clearly an unphysical result, because the evolution map acts locally, 
{\it i.e.} in a separate form, and therefore can not create quantum correlations.
The existence of such states implies that the ``slippage prescription'' should 
take care not only of single system states developing negative eigenvalues, 
but also of possible inconsistencies related to the entanglement
of these systems with any other ancilla. 

In order to explicitly expose this inconsistency, 
it will be sufficient to work within a special class of 
two-qubit density matrices, those for which the non-vanishing entries lie
along the two diagonals:
\begin{equation}
\label{16}
\rho=\pmatrix{ 
\rho_{11} & 0 & 0 & \rho_{14} \cr
0 & \rho_{22} & \rho_{23} & 0 \cr
0 & \bar{\rho}_{23} & \rho_{33} & 0 \cr
\bar{\rho}_{14} & 0 & 0 & \rho_{44} 
}\ .
\end{equation}
Further restrictions on the entries
of this matrix need to be imposed in order to represent a state.
In particular, the trace must be one, 
$\rho_{11}+\rho_{22}+\rho_{33}+\rho_{44}=1$,
while the positivity of the
spectrum implies the positivity of the two subdeterminants
\mbox{$\rho_{11}\rho_{44}-|\rho_{14}|^{2}$} 
and \mbox{$\rho_{22}\rho_{33}-|\rho_{23}|^{2}$} and of the entries along the diagonal.
The form (\ref{16}) is particularly suited for our considerations since it is 
preserved by the action of the dynamics
$\gamma_t\otimes{\rm id}$; further, its entanglement content can be 
explicitly calculated.

In this respect, a convenient measure of entanglement
is provided by concurrence:
\begin{equation}
{\cal C}(\rho)=\max\{0,R_{1}-R_{2}-R_{3}-R_{4}\}\ ,
\label{17}
\end{equation}
where $R_{i}$ are the square roots of the eigenvalues of 
$R=\rho\ (\sigma_{2}\otimes\sigma_{2})\ \rho^{*}\ (\sigma_{2}\otimes\sigma_{2})$ taken in decreasing order;
it vanishes for a separable state while takes positive 
values between zero and one for entangled states [21, 22].
For the state (\ref{16}), one explicitly finds:
\begin{equation}
{\cal C}(\rho)=\max\{0,\max\{|\rho_{23}|-\sqrt{\rho_{11}\rho_{44}},|\rho_{14}|
-\sqrt{\rho_{22}\rho_{33}}\}\}\ .
\label{18}
\end{equation}
It is then clear that the state (\ref{16}) is entangled provided 
$\max\{|\rho_{23}|-\sqrt{\rho_{11}\rho_{44}}$,\break$|\rho_{14}|-\sqrt{\rho_{22}\rho_{33}}\}>0$.
For simplicity, in the following we shall assume to start at $t=\,0$ with an entangled state $\rho(0)$
fulfilling the more restrictive condition 
$|\rho_{23}|-\sqrt{\rho_{11}\rho_{44}}>|\rho_{14}|-\sqrt{\rho_{22}\rho_{33}}>0$.

Let us then consider the following two-parameter family of states
\begin{equation}
\label{19}
\hskip -1cm
\rho=\pmatrix{ 
\mu & 0 & 0 & -\frac{a}{b}\nu \cr
0 & \frac{1}{2}\vartheta(1-3\mu)+\frac{1}{2}(1-2\mu) & i\nu & 0 \cr
0 & -i\nu & \frac{1}{2}\vartheta(3\mu-1)+\frac{1}{2}(1-2\mu) & 0 \cr
-\frac{a}{b}\nu & 0 & 0 & \mu}\ .
\end{equation}
where $\mu$ and $\nu$ are real constants satisfying the three constraints
(necessary for positivity)
\begin{eqnarray}
\label{20}
\nonumber
\\
\nonumber
\frac{1-\vartheta}{3-2\vartheta}<\mu<\frac{1+\vartheta}{3+2\vartheta}\ , \\
\nonumber
\\
\frac{-2+3\vartheta^{2}-\sqrt{4-3\vartheta^{2}}}{9\vartheta^{2}}<\mu
<\frac{-2+3\vartheta^{2}+\sqrt{4-3\vartheta^{2}}}{9\vartheta^{2}}\ , \\
\nonumber
\\
\nonumber
\frac{1}{2}\sqrt{(1-2\mu)^{2}-\vartheta^{2}(3\mu-1)^{2}}>\nu>\mu>\frac{a}{b}\nu\ ,
\nonumber
\\
\nonumber
\end{eqnarray}
with $\vartheta=d/a$. 
In writing (\ref{20}), we have assumed $a<b$; this is not really restrictive 
since for $a> b$ a similar family of states can be found.
One can check that these matrices represent initially entangled two-qubit states that 
remain positive under the evolution $\gamma_t\otimes{\rm id}$; therefore, they
are admissible states within the ``slippage prescription''.%
\footnote{The system of inequalities (\ref{20}) has solutions only when $\vartheta$
takes values in a certain range, which, recalling the condition (\ref{10}),
is related to the temperature of the bath. In our discussion we have taken
$\sqrt{3}/2\leq\vartheta\leq1$, since this allows certain simplifications in the
calculations.} 

However, these states present another, more subtle inconsistency than non-positivity.
Indeed, using numerical methods, one can show that their concurrence increases for small times.
The picture below displays the behaviour of the concurrence 
of one of these states as a function of time.%
\footnote{The graph is drawn for the following representative values of the basic
parameters: $a/\omega=0.007$, $b/\omega=0.01$, $d/\omega=0.0065$, with time in
units of $1/\omega$.}
It shows an oscillatory behaviour that is in clear contradiction with quantum mechanics, since
the dynamics is in factorized form.

\begin{figure}[h!]
\begin{center}
\includegraphics{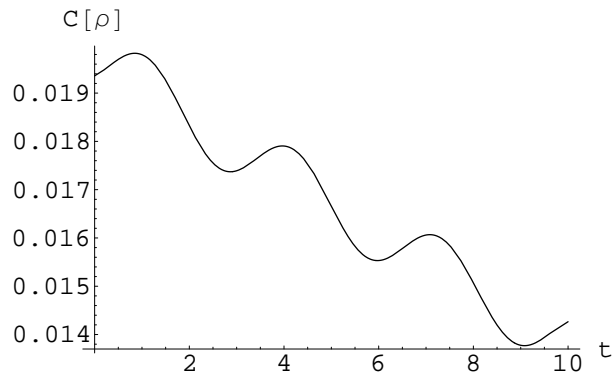}
\caption{Concurrence behavior in time}
\end{center}
\end{figure}

This unphysical behaviour of the concurrence can be studied
analytically in the case of zero temperature;
in fact, recalling (\ref{10}), $\beta^{-1}=\,0$ implies the simplifying condition $d=a$.
In this case, the matrix expression of our states and the corresponding constraints
on the parameters $\mu$ and $\nu$ reduce to:
\begin{equation}
\rho=\pmatrix{
\mu & 0 & 0 & -\frac{a}{b}\nu \cr
0 & 1-\frac{5}{2}\mu & i\nu & 0 \cr
0 & -i\nu & \frac{\mu}{2} & 0 \cr
-\frac{a}{b}\nu & 0 & 0 & \mu}\ ,
\label{21}
\end{equation}
and
\begin{equation}
\label{22}
0<\mu<\frac{2}{9}\ ,\qquad
\frac{1}{2}\sqrt{2\mu-5\mu^{2}}>\nu>\mu>\frac{a}{b}\nu\ .
\end{equation}
With these conditions, it is easy to verify that $\rho\geq0$. 

The evolution in time of this matrix under $\gamma_t\otimes{\rm id}$ can be obtained 
from (\ref{12}); using 
the labelling introduced in (\ref{16}) for the entries of $\rho$, 
one explicitly finds:\hfill\break
\begin{eqnarray}
\hskip -2cm
\rho_{11}(t)=e^{-2at}\mu\ , \\
\hskip -2cm
\rho_{22}(t)=1-\frac{3\mu}{2}-\mu e^{-2at}\ , \\
\hskip -2cm
\rho_{33}(t)=\frac{\mu}{2}e^{-2at}\ , \\
\hskip -2cm
\rho_{44}(t)=\frac{3\mu}{2}-\frac{\mu}{2}e^{-2at}\ , \\
\hskip -2cm
\rho_{14}(t)=\overline{\rho_{41}(t)}=e^{-at}\left[-\frac{a\nu}{b}\cos(\Omega t)
-\frac{b\nu}{\Omega}\sin(\Omega t)+i\frac{a\nu}{b\Omega}\sin(\Omega t)(\omega+2b)\right]\ , \\
\hskip -2cm
\rho_{23}(t)=\overline{\rho_{23}(t)}=e^{-at}\left[\frac{\nu}{\Omega}\sin(\Omega t)\left(-\frac{a^{2}}{b}-\omega-b\right)+i\nu\left(\cos(\Omega t)+\frac{a}{\Omega}\sin(\Omega t)\right)\right]\ .
\label{28}
\end{eqnarray}

As previously mentioned, 
the positivity of the state $\rho(t)$ at time $t$ is assured by the positivity 
of the two sub-determinants 
\mbox{$\rho_{11}(t)\rho_{44}(t)-|\rho_{14}(t)|^{2}$} 
and \mbox{$\rho_{22}(t)\rho_{33}(t)-|\rho_{23}(t)|^{2}$}; in this case, these conditions read
\begin{eqnarray}
\hskip -2.5cm
\frac{\mu^{2}}{2} (3-e^{-2at})-\nu^{2}\left[\left(\frac{a}{b}\cos(\Omega t)
+b\frac{\sin(\Omega t)}{\Omega}\right)^{2}
+a^{2}\frac{\sin^{2}(\Omega t)}{\Omega^{2}}\bigg(2+\frac{\omega}{b}\bigg)^{2}\right]\ge0\ , \\
\hskip -2.5cm
\frac{\mu}{2}(1-\frac{3}{2}\mu-\mu e^{-2at})
-\nu^{2}\left[\frac{\sin^{2}(\Omega t)}{\Omega^{2}}\bigg(\frac{a^{2}}{b}
+\omega+b\bigg)^{2}+\left(\cos(\Omega t)+a\frac{\sin(\Omega t)}{\Omega}\right)^{2}\right]\ge0\ .
\label{30}
\end{eqnarray}
In order to verify that these inequalities are indeed satisfied, 
recall from (\ref{8}) that $a$, $b$, $d$ are proportional to $\lambda^{2}$;
since $\lambda$ is by assumption small, one can take
$a,b\ll\omega$; being also $\Omega^{2}=\omega^{2}+2\omega b-a^{2}\sim\omega^{2}$, 
we can neglect $a$ and $b$ with respect to $\omega$ and $\Omega$ 
and then discard the terms proportional to $a/\Omega$, $b/\Omega$ and their powers
with respect to those proportional to $a/b$ or $\omega/\Omega\sim1$.
As a consequence, the conditions (29), (30) reduce to
\begin{eqnarray}
\mu^{2}\big(3-e^{-2at}\big)\ge	\frac{2 a^{2}}{b^{2}}\nu^{2}\ , \\
\mu\bigg(1-\frac{3}{2}\mu-\mu e^{-2at}\bigg)\ge	2\nu^{2}\ ;
\label{32}
\end{eqnarray}
these are easily seen to be satisfied thanks to the constraints in (\ref{22}).
In conclusion, the density matrices in (\ref{21}) are admissible initial states for the non-positive evolution
$\gamma_t\otimes{\rm id}$, since they remain positive for all times.

Let us now compute their concurrence; one explicitly finds:
\begin{eqnarray}
\hskip -1cm
{\cal C}(\rho(t))=&&\nu e^{-at}\sqrt{\left(\frac{a^{2}}{b}
+\omega+b\right)^{2}\frac{\sin^{2}(\Omega t)}{\Omega^{2}}
+\left(\cos(\Omega t)+\frac{a}{\Omega}\sin(\Omega t)\right)^{2}}\\
&&\hskip 7cm -\frac{\mu}{2}e^{-at}	\sqrt{6-2e^{-2at}}\ .
\nonumber
\end{eqnarray}
It is sufficient to examine the behaviour of $\cal C$ for small times:
\begin{equation}
{\cal C}({\rho}(t))\simeq\nu-\mu+\frac{a\mu}{2}t+O(t^{2})\ .
\end{equation}
Since $a$ is positive, from this expression one immediately concludes
that indeed ${\cal C}({\rho}(t))$ increases in time.

\section{Discussion}

It is widely believed that the dynamics of a subsystem in weak interaction with an external environment
can be described in terms of semigroups of linear maps $\gamma_t$
generated by a Markovian master equation. In order to be physically acceptable,
this effective description needs to
satisfy basic physical requirements.
In the first place, it must preserve the 
positivity of any initial density matrix, since their eigenvelues
represent probabilities.
In the second place, one has also to care of possible couplings with
another system, not subjected to noise and inert, and therefore to guarantee the positivity-preserving 
character also of the semigroup of maps of the form $\gamma_t\otimes {\rm id}$,
as studied in the previous section.

Unfortunately, most phenomenological derivation of reduced dissipative
dynamics lead to semigroups of linear transformation that are
not positive. 
To avoid inconsistencies, one usually restricts the possible initial states 
to those for which $\gamma_t$ remains positive 
(the so-called ``slippage of initial conditions''). 
This prescription works also in the case of the evolution $\gamma_t\otimes {\rm id}$
for two subsystems, 
provided the initial state is in separable form:
$\rho(0)=\sum_i p_i\, \rho^{(1)}_i\otimes\rho^{(2)}_i$,
$p_i\geq0$, $\sum_i p_i=1$, where $\rho^{(1)}_i$ and $\rho^{(2)}_i$
are admissible states for the first and second subsystems, respectively.

However, as shown in the previous section, when the initial state 
$\rho(0)$ is not in factorized form but still remains positive under
the action of the non-positive dynamics $\gamma_t\otimes {\rm id}$, further,
more subtle inconsistencies may arise.
Indeed, we have found explicit examples of two-qubit states which 
under the action of $\gamma_t\otimes{\rm id}$ present an increasing concurrence. 
The creation of entanglement by a local operation is clearly unacceptable on physical grounds. 
This means that the ``slipped'' dynamics is still not free from inconsistencies.

As a consequence, in order to continue to use non-positive reduced dynamics of Redfield type,
a new, more general ``slippage'' mechanism should be invoked:
it must take care not only of states developing negative eigenvalues but also of those
presenting unphysical increase in entanglement.
The only way to practically implement it is by further restricting
the space of initial admissible states, discarding also some entangled ones. 

It should be noticed that possible inconsistencies are not limited to the two-qubits case;
by considering more complicated ancillary coupling similar problems may arise for
multipartite entangled states that should therefore also be eliminated by the ``slippage 
operation''. The risk of such a mechanism is to restrict too much the space of states, 
losing, in particular, many entangled states.
These considerations seems to suggest that there is an intrinsic
incompatibility between the existence of entangled states
and the slippage prescription adopted to cure the
inconsistencies that non-positive, reduced dynamics might produce.

In closing, let us mention that in the few cases for which the Markovian limit of the subdynamics can be obtained in a rigorous way, the resulting evolution map
$\gamma_t$ turns out to be not only positive, 
but also completely positive [1-6]. 
In these cases, the compound map $\gamma_t\otimes{\rm id}$ is also completely positive and therefore no inconsistencies can arise, even when acting on entangled states.

\end{document}